# Polyelectrolyte intelligent gels. Design and applications

**Piero Chiarelli\* and Danilo De Rossi \*\***

**\* CNR…**

**\*\* Interdepartmental Research Centre "E.Piaggio"- University of Pisa**

# CHAPTER 15

**Polyelectrolyte intelligent gels. Design and applications**

Piero Chiarelli and Danilo De Rossi

## 15.1 Introduction and technical overview

Synthetic, biological and hybrid hydrogels have found a large variety of uses, mostly related to biomedical applications[1]. Stimulus-responsive polyelectrolyte gels represent a niche area whose interest is growing; these environmentally responsive gels are often referred to as smart or intelligent gels[2].

Reversible swelling of partially ionized polyelectrolyte gels mediated by external stimuli has attracted considerable interest by scientists and engineers in the last 60 years[3].

The field has essentially started with the independent work of J.W. Braitenbach[4], W. Kuhn[5] and A. Katchalsky [6], all published in 1949, reporting how reversible titration of weakly ionized polyelectrolyte gels provides conspicuous swelling and deswelling of the polymer network. These works inaugurated the field of "polymer mechanochemistry" or, as redefined later, "polymer chemomechanics"[7].

In more recent years, theoretical work[8] and experimental observations of discrete phase transitions in ionic gels under appropriate conditions[9] were reported.

These observations sparkled active research in the field. Many other external stimuli beyond pH changes proved to be effective in generating macroscopic volume changes in polyelectrolyte gels. These changes are also accompanied by large variations in physico-chemical properties of the gels. Temperature, ion or solvent exchange, electric and magnetic fields, radiant energy and also biochemical reactions or immobilised ligand interactions have all been demonstrated to be effective external stimuli on specifically tailored gels [10]. Several areas of application are thought to greatly benefit of smart properties of polymer gels and a large amount of research work is nowadays devoted to the development of materials and devices which could exploit stimulus-responsive behaviour, particularly in the biomedical field[11][12][13][14].

Early work on stimuli-responsive gels was largely focused on the implementation of macroscopic muscle-like actuators driven by chemicals[15], by electric fields [16] or by electrochemical reactions [17]. More recent work has addressed the development of drug delivery systems [18], micro fluidic components and circuits [19], sensors [20], biosensors [21], optical components [22], active membranes for separation [23], smart surfaces [24], and scaffolds for tissue engineering [25].

It is now evident that research efforts in the field of macromolecular engineering has led to substantial advances to the endeavour of mimicking through synthetic pathways some functional characteristics of biological engines.

While applications in tissue engineering, active surfaces and drug delivery systems typically do not require gels to be mechanically strong, other applications, in particular muscle-like actuators, need tough materials, fast swelling-deswelling response and the capability to undergo many cycles without damage and with minimum hysteresis. Stimulus-responsive gels endowed with these optimized properties are not yet available and devices which have been proposed are just proofs of the difficulties involved.

Strong hydrogels have been recently obtained [26]. These gels might open new avenues in more demanding applications; however the intrinsic biphasic (liquid-solid) nature of gels makes necessary most devices being encapsulated, adding additional complexity to design which engineers could only accept in view of unique performances.

In this chapter polyelectrolyte intelligent gels are examined along three broad lines. The effects of different physical, chemical and biological stimuli on gels response are analysed and mechanisms of response are outlined. The broad range of biomedical applications of smart gels is reviewed and limits and perspectives of the proposed techniques and devices are critittically discussed. Finally, continuous modelling of gel electromechanochemistry is described, providing quantitative tools to assess swelling equilibrium conditions and coupled kinetics.

## 15.2 Models of Swelling Equilibrium and Kinetics

Modelling the response of a polyelectrolyte gels to stimuli of different nature to quantify their swelling equilibrium and kinetic behaviour is a complex task and it can be attempted at different scales and very different levels of approximation[27]. The following section (15.2.1) summarizes the thermodynamic approach to the equilibrium behavior of gels. Modelling with the intent of designing devices and predicting their response is best accomplished by recurring to macroscopic, phenomenological continuum or discrete description of the solvent-network system. Continuum models of gel electromechanochemistry are now available in the literature[28] as much as analytical solutions for simple geometries and strong approximations and numerical solution of more rich formulations and descriptions. We summarize in the Appendices the basic concepts and derivations which consent to grasp the essentials of the modeling efforts which have been undertaken.

### *15.2.1 Gel bi-phasic systems: free energy – phase diagrams*

A gel consists of a solid polymeric network and a liquid phase that fills its interstices. The polymeric matrix may have ionizable groups on its backbone with small ions of opposite charge (counter-ions) dissolved in the interstitial fluid to maintain electroneutrality.

The free energy steaming out by the solid-fluid interactions prevents the polymer network to collapse, conferring to polyelectrolyte gels their characteristic shape, volume, elasticity, porosity etc. All these equilibrium properties depend on physical and chemical parameters such as temperature, pH, ionic strength of the solution, type of solvent.

The studies of Flory and Huggins [29-30] have put into evidence that a gel can be schematized as a physical system governed by osmotic forces which define its volume, shape, elasticity. These osmotic forces $\Pi_i$ are defined by how free energy $\Delta G_{tot} = \sum_i \Delta G_i$ changes in respect to variation of the gel extensive variables such as its volume following the rule

$$\Pi_i = -\partial(\Delta G_i)/\partial V \tag{1}$$

The most important forces that generally operate in gels originate from rubber elasticity and polymer-solvent affinity, while those deriving from the counter-ion osmotic pressure and electrostatic repulsion (or attraction) on the polymer chains are present in charged gels.

The polymer-solvent affinity quantifies how the polymer-solvent contacts are energetically preferred with respect to the polymer-polymer ones. Greater is the polymer solvent-affinity, higher is the osmotic pressure that sucks the fluid into the matrix network. This part of the free energy $\Delta G_m$ depends upon the polar character of the solvent molecules and from the hydrophobic-hydrophilic character of the groups linked to the polymer backbone following the Flory-Huggins formula[29]

$$\Delta G_m = RT\,[n_1 \ln v_1 + n_2 \ln v_2 + n_1 v_2 \chi_{12}] \tag{2}$$

where $n_1$ and $n_2$ are the moles of solvent and polymer respectively, and $v_1$ and $v_2$ are their respective volume fractions; $\chi_{12}$ is a "dilution" parameter that takes into account the increase of energy due to the polymer-solvent contacts. From (2) the osmotic pressure of polymer-solvent mixing reads

$$\Delta\Pi_m = -(V_1)^{-1} \partial(\Delta G_m)/\partial n_1 = (V_1)^{-1} RT [\ln(1-v_2) + v_2 + v_2 \chi_{12}] \qquad (3)$$

where the identity $v_1 + v_2 = 1$ has been used, and $V_1$ is the molar volume of solvent.

The rubber elasticity of the polymer network comes out from the tendency of each statistical coil to maintain its end-to-end distance between two network cross-links. Rigid units that are free to rotate with respect to the axis of the preceding one generate the statistical coil of the polymeric chain. Thermal fluctuations make these units to change continuously their reciprocal angles letting the end-to-end length of the coil to fluctuate around a defined mean distance. Changing the intensity of the thermal fluctuations, the mean end-to-end distance of the polymeric coil will vary. Moreover, higher is the temperature stronger is the tendency of the polymeric coil to maintain its mean undeformed end-to-end distance, leading to an increase of the macroscopic network elastic moduli. This behavior is described by the free energy of the polymer network that reads[29]

$$\Delta G_r \cong \tfrac{1}{2} n_v RT \{\alpha_x^2 + \alpha_y^2 + \alpha_z^2 - 3 - \ln[\alpha_x \alpha_y \alpha_z]\} \qquad (4)$$

where $n_v$ is the number of moles of polymer chains, $\alpha_i = r_i / r_{0i}$, $r = (r_x, r_y, r_z)$ is the end-to-end distance and $r_0$ is the reference (isotropic, undeformed) length. In the affine deformation approximation, the chain deformation parameters correspond to the macroscopic strain of the network.

For isotropic swelling (i.e., $(\alpha_x \alpha_y \alpha_z) = V/V_0$, $\alpha_x^2 + \alpha_y^2 + \alpha_z^2 = 3(V/V_0)^{2/3}$) the expression (4) leads to the osmotic pressure of the rubber elasticity that reads

$$\Delta\Pi_r = -(V_1)^{-1} \partial(\Delta G_r)/\partial n_1 = -\tfrac{1}{2} n_v RT \, \partial\{3(V/V_0)^{2/3} - \ln[V/V_0]\}/\partial V$$

$$= -n_v RT \{(V_0/V)^{1/3} - \tfrac{1}{2}(V_0/V)\}/V_0^3 \qquad (5)$$

where V is the gel volume and $V_0 = r_{0x} r_{0y} r_{0z}$. For polyelectrolytes that undergo large swelling (i.e., $V \gg V_0$ and $(V/V_0)^{1/3} \ll V/V_0$) it holds

$$\Delta\Pi_r \cong -n_v RT / V^{1/3} V_0^{2/3} \qquad (6)$$

The electrostatic repulsion of the charges fixed to the polymer coil leads to positive gel swelling pressure. This term can be calculated by means of the Debye-Huckel theory applied to the polyelectrolyte chains. Even if the majority of the counter-ions are captured by the ionic atmosphere around the charged polyelectrolyte chains, two contributions to the gel free energy remain: the electrostatic interaction between the fixed charges and the osmotic pressure of the free counter-ions that cannot leave the gel network because of the Donnan potential. The first contribution[30] (that is influenced by the dielectric constant of the interstitial solvent, its ionic strength and pH) is smaller than the other ones and is usually disregarded, particularly when a salt is added to the swelling solvent.

At equilibrium the diffusional (chemical) force of the free counter-ions to leave the gel is balanced by the electrostatic force so that the total force as well as the electrochemical gradient at the gel-external bath is null. The Donnan electrical potential $\Delta \Psi$ equals the chemical one following the formula

$$\Delta \Psi = \pm RT \ln[C_i^\pm / C^*{}_i^\pm]/F \qquad \forall i \qquad (7)$$

where $C_i^\pm$ and $C^*{}_i^\pm$ are the anionic and cationic concentrations inside the gel and in the external bath, respectively. The relation (7) fixes the ratio of each ionic species inside and outside the gel so that with the mass conservation defines the ionic concentrations of the gel system. The Donnan equilibrium in the presence of an added salt requires a disproportion of individual mobile ions within the gel and in the equilibrium excess solution. For instance, for a cationic polymer of fixed charge Z, there will be more mobile anions in the solution within the gel ($m_-$) than in the excess solution ($m'_-$). The disproportion decreases with the overall salt concentration, and the disproportion coefficients ($m_- / m'_- = m'_+ / m_+$) attain the value of 1 above the electrolyte concentration of 0.2 M, when screening of the polymer charge is almost complete (see Chapters 3 and 6).

The osmotic force of the free counter ions, which are confined inside the gel, exerts a positive pressure that inflates it like a gas confined inside a container. This contribution increases with temperature and with the dissociation of the ionizable groups along the polymer chain. Dissociation, on its turn, is influenced by the pH and ionic strength of the interstitial solution. Given the Donnan potential (7) the free energy and osmotic pressure read

$$\Delta G_{os}= \alpha_i \, n_i \, F\Delta \Psi = - \sum_i \alpha_i^2 \{n_i \, RT \ln [n_i / V] + n^*_i \, RT \ln [n^*_i / V^*]\} \quad (8)$$

$$\Delta \Pi_{os}= - (V_1)^{-1} \partial(\Delta G_{os})/ \partial n_1 = \sum_i n_i \, RT / V - \sum_i n^*_i \, RT / V^* \quad (9)$$

where $\alpha_i = q_i/|q_i|$, $q_i$ is the charge of the i-th ionic specie, $n^*_i$ are the i-th ionic concentration and $V^* = V_{tot} - V$ the volume of the external bath in equilibrium with the gel. The expression (9) is quite complicate since the disproportion between $n_i$ and $n^*_i$ as well as the effective charge on polymer chains are functions of the gel volume.

Nevertheless, it is interesting to derive the osmotic pressure in a simple case. If we suppose that the external bath is infinitely larger than the gel (i.e., $V^* >> V$ and hence $V_{tot} >> V$) and that the gel is electro-neutral (we disregard infinitesimal amount of ions needed to generate the Donnan potential), $C^*_i{}^\pm$ as well as the effective fixed polymer charge at different gel volume can be considered practically constant. Thence, in absence of support electrolyte, so that the moles of counter-ions $n_c$ inside the gel are constant, (9) leads to the expression

$$\Delta \Pi_{os} \cong n_c \, RT / V + \Pi_{ext} \quad (10)$$

$$\Pi_{ext} = - n^*_c \, RT / (V_{tot} - V) \cong - C^*_c \, RT = \text{constant} \quad (11)$$

A compact expression for the degree of swelling of a polyelectrolyte network as a function of the degree of cross-linking (elastic contribution), the interaction parameter (mixing contribution), the fixed charge and ionic strength (ionic contribution) was earlier given by Flory[29]

This plurality of energetic contributions, having no equals in other non-biological material systems, is able to confer to gels a complex behavior sensible to many physico-chemical inputs.

The counter-ions osmotic pressure is usually positive and tends to inflate the gel, while the contribution of rubber elasticity is typically negative and it counterbalances the small ions osmotic pressure giving to the gel volume a stable condition. The others terms can be either positive or negative, displacing the equilibrium point.

Generally speaking, the isothermal curves in the gel pressure-volume phase diagram have a hyperbolic decreasing shape. If we try to increase the gel volume (at zero total pressure) the positive osmotic pressure of the small ions decreases faster than the negative rubber elasticity, so that the total gel pressure needed to maintain this new state is lower and of negative value.

Since the rate of change of each osmotic pressure term is not linear, we may have the appearance of a flex in the phase diagram with a critical point as well as unstable states with the typical bell-shaped area.

In this domain of states, for instance, the rate of decrease of the small ions pressure may be slower than the rate of increase of the polymer-solvent affinity and, therefore, a volume increases leads to an increase of total gel pressure, generating a further expansion. As it is well known, this kind of instability is associated to a phase transitions. In this case, the gel makes an abrupt large step expansion to a new equilibrium volume[31].

A continuum kinetic model for the network-solvent readjustment dynamics, induced e.g. by a phase transition, is outlined in the Appendix.

## 15.3 Physically responsive gels

Behind similar kinetic of the gel network readjustment, there may be very different phenomena on molecular scale activated by the external stimulus.

In this section we are going to examine the variety of the actuation mechanisms in gel systems grouped by the typology of the stimulus.

### *15.3.1 Thermally sensitive gels*

Temperature sensitive gels are based upon the principle that a polymer coil may undergo conformational changes with temperature eventually resulting in gel volume changes. This process is induced by the change of the polymer-polymer contact energy with respect to the polymer-solvent one when the temperature is changed.

Poly-N-isopropyllacrylamide (NIPAM) gels constitute one of the most investigated systems. These supermolecular aggregates transduce on a macroscopic scale the conformational changes that happen on a molecular one where polymer shrinks as temperature increases.

In a solution the polymer solute separates into polymer-rich and diluted phase as the lower critical solution temperature (LCST) is reached[34]. When the polymer coils are cross-linked the resulting gel undergoes volume deswelling at a temperature that is close to the one of the polymeric solution.

A swollen NIPAM gel undergoes shrinking at 40°C[35]. Homopolymer gels have a sharp volume collapse that can be recognized as true phase transitions, while functionalized copolymer with, for instance, ionizable acrylic acid groups[36] have a more smooth gel deswelling at the critical temperature.

The gel readjustment kinetics and the thermal diffusion in these gels are coupled each other. The simplest behavior happens in macro-porous gels where the thin gel pore wall mechanically responds very quickly to the temperature change.

In macro-porous gels, hence, the limiting rate phenomenon is the heat diffusion into the macro-pores. In the gel swelling process (temperature decrease below the LCST) the solvent uptake by the gel enhances, by convection, heat transfer to the inner part of the network. In the inverse process (gel deswelling) the heat transfer is hindered by the out-flowing of the solvent from the inside of the polymer matrix thus generating an asymmetry (hysteresis) in the contraction-elongation cycle.

In homogeneous gel the heat transfer inside the gel can only occur by thermal diffusion since the water uptake or release, due to the mechanical readjustment, is very slow. Moreover, given that the mechanical parameters depend also by the state of the gel, in NIPAM gel with a large gel deswelling, the solvent diffusion through the collapsed domains can be lowered practically to zero.

When the temperature change is induced by the gel immersion in an external bath, the collapsed domains may constitute an external skin around the swollen gel and can prevent the outflow of fluid from inside the gel and the consequent mechanical readjustment.

Others gels which also show thermal response are the polyvinyl-methyl-ether[37] which progressively undergoes swelling below the temperature of about 37 °C, poly-N-vynilcaprolactam/ethilene glycol and dimethacrylate/divinyl sulfone[38] which has a glassy-like collapsed state above the LCST, and hydroxypropylcellulose [39-40] which has the peculiarity to decrease its elastic modulus in the deswelled state.

Even if the thermal response is highly reversible, the major obstacle to the use of these kinds of gel actuators (for instance in the field of drug delivery) is the very low cooling rate obtainable by spontaneous heat diffusion.

### 15.3.2 Electro-magneto sensitive gels

The responsivity of polyelectrolyte gels in solution to electric fields is governed by different phenomena whose importance depends upon several factors such as potential difference at the electrodes, geometry and others.

Even from the very first experimental observation of gels responsivity to electrical stimuli, diverse mechanisms have been postulated to operate, although many, if not all of them, are concomitantly active[41].

Hamlen et al.[17] and subsequently Fragala et al.[42] exploited electrolytic reactions or protons electrodialysis to swell and shrink pH sensitive polyelectrolyte gels firstly exploiting electrical stimuli.

Tanaka [43] reported progressive collapse of a weakly ionized polycationic gel in contact with a metal anode, immersed in acetone-water (50:50) mixture under an electrode potential difference ranging from 1.25 to 2.15 v. Gel deswelling was ascribed to the electrophoretic attraction of the charged gel network to the electrode resulting in the generation of a mechanical stress gradient (orthogonal to the electrode) squeezing the gel.

Osada and Hasebe[44] reported pAMPS and other gels shrinking and water exudation when in contact with carbon electrodes under DC excitation.

The phenomena were tentatively ascribed to concomitant electrophoretic attraction of the gel to the electrode and electrostatically induced gel dehydration. De Rossi et al.[45], working with PAA-PVA gels in solution with slightly higher potential difference at the electrodes observed gel swelling or shrinking governed by water electrolysis at interfaces, resulting in local pH changes.

Gel bending under the action of an electric field was first reported by Shiga and Karauchi [46] under appropriate conditions and geometry.

Alternate bending of gels leading to a worm-like motility was reported by Osada et al.[47] by exploiting selective charged surfactant binding driven by an electric field and causing osmotic pressure changes at the gel surface resulting in mechanical actions.

The electrochemical response of gels to copper oxidation to Cu2+ has also been described [48]. In this case the double interaction of each ion Cu2+ with the anionic ionizable groups on the polymer network generates additional cross-links increasing its rubber elasticity and leading to de-swelling as the Cu2+ concentration increases.

Gels have been prepared containing magnetic particles in their soft matrices that react to magnetic fields[49] as well as to electrical ones[50] by changing their shape and physical properties similarly to as magneto-rheological and electro-rheological fluids.

Gels of silicone rubber or oil, and polyurethanes with dispersed micron-sized particles iron have been synthesized[51]. Such gels respond to the magnetic field by increasing their stiffness, so that an increase of stress can be obtained by a pre-strained sample.

A carrageenan hydrogel with barium ferrite micro-particles showed a decrease of its elasticity when subject to a magnetic field[52].

### 15.3.3 Light sensitive gels

Light can also be used to swell and deswell gels. One possibility relies upon the indirect stimulation of thermally sensitive gels by absorption of radiant energy to increase their temperature.

One example is given by Suzuki and co-workers that synthesized a NIPAM gel containing copper chlorophyllin that swells and contracts as light is turned on and off [53].

A similar response was also obtained in NIPAM gels with dispersed nanoparticles of gold[54] that respond to near-IR electromagnetic radiation[55].

A different mechanism leading to light sensitive gels is constituted by the use of light-induced conformational changes such as the photo-isomerization of azobenzene between the *cis* and *trans* forms[56] that changes the rigidity of the polymer network under UV irradiation. Ionization of leucocyanides groups under UV that can lead to osmotic swelling of the hosting gels[57] and nitrocinnamate groups that reversibly undergoes cross-linking and cleavage under photo irradiation have also been reported[58]. The *keto* to *enol* tautomerization was also shown to be able to induce gel shape change under UV irradiation[59].

Polyacrylic acid gels cross-linked by means of copper ions and containing titanium dioxide have shown undergoing swelling due to the pH change induced by UV light[60]. Other similar gel systems have been synthesized by making use of silver-coated titanium dioxide microparticles.

**15.4 Chemically responsive gels**

*15.4.1 pH and salt sensitive gels*

Since the pioneering works of Kunh and Katchalski in the fifties, gels where investigated as chemo-mechanical systems stimulated by means of monovalent and bivalent cations. The muscle-like contraction-elongation behavior has attracted the attention of many researchers since nowadays, even if the biological analog is much more complex and it exploits a very different contractile mechanism.

The most investigated hydrogel is cross-linked polyacrylic acid. The solubility of the polymer in water as a function of pH almost defines the state of the gel network in the macromolecular aggregation. At low pH the ionizable groups are not charged so that the electrostatic repulsion is almost absent as well as the counter-ions responsible for the positive osmotic pressure. In this state the gel is in the shrunken state.

At high pH (>5), much above the carboxylic groups pKa, the functional groups are ionized with counter-ions that warrant electro-neutrality. In this salt form the gels adsorb water and remain in the swollen state.

As shown in section **Appendix** the rate of the gel volume change depends by its geometrical form. A polymer that can be shaped in form of very thin fibers can undergo a rapid contraction-elongation cycle (1-10 seconds). Moreover, the anisotropic orientation of the networks chains can increase the elastic modulus of the gel and its stress generation along a preferred direction in response to pH stimuli. This is the case of polyacrylonitrile fibers that under thermal treatment in basic conditions can be converted in ionized gel fibers [61].

Alternatively, the time response can be shortened by synthesizing a gel with a macro-porous structure by means of freezing-thowing[62] or by freezing drying cycles [63].

Others polyacid gels that react to pH and salt concentration variations have been synthesized [64-65] such as those based on polymethacrylic acid [7]. If the gel network owns strong acid groups, such as the sulfonated one, the swelling is translated to lower pH as well the pKa value.

Other gels, such as phosphatated ones, can undergo multiple ionization process where the functional groups progressively reach the dissociated form.

Symmetrically to polyacids, polybasic gels shrink in basic conditions and swell in acid ones. In this case, the functional groups, such as that one of the aminomethylmethacrylate monomer, lead to polybasic gels. Another type of polybase is given by the epoxy-amino gels[66].

A polymeric network owning both acidic and basic side groups has anphoteric properties showing a more complex swelling behavior [67-68]. These gels also react to changes of salt concentration in the interstitial solution as a consequence of the shielding of the charges present onto the matrix by the salt ions.

*15.4.2 Chemical reaction sensitive gels*

The number of gels responding to chemical, or biochemical reactions and non-linear covalent cooperative binding[69] is rapidly increasing due to the perspective of innovative applications in the biosensors and drug delivery fields.

In this paragraph we firstly consider those gels that are sensible to simple chemical reactions leaving the others to the next sections.

An interesting class of gels concerns those ones that swell and contract cyclically in response to oscillating chemical reactions.

A co-polymer of NIPAM gel has shown[70-71-72] to cyclically contract and swell following the oscillating Belousov-Zhabotinsky reaction.
The mechanical oscillation is generated by the oxidation and reduction of the ruthenium(II) tris(2,2'-bipyridine) group, covalently bonded to the polymer chain, which autonomously periodically switch between the two states in a closed solution without any external action.

Also pH oscillations around acidic block copolymer have shown to induce cyclic swelling and shrinking in gels[73-74].

Even if there exists a formal analogy with the biological contraction, the energy densities involved in those systems are much smaller than those of natural muscles.

Gel systems that respond to a chemical input via covalent bond formation to the analyte have been recently developed for the detection of amines, alcohols, aldehydes, carbon dioxide, saccharides, thiols, hydrogen sulfite, hydrogen disulfide, cyanide and amino acids[75].

The glucose-sensitive gels via boronic acids[76] are the most investigated since enzymes-based hydrogels, raised concerns about their stability, toxicity and undesirable immunogenic responses.

A gel has been synthesized via the free radical polymerization of N-vinyl-2-pyrrlidone and m-acrylamidophenylboronic acid. A polymeric complex was then formed via boronate ester formation to the diol units of polyvinyl alcohol. Viscosity was shown to markedly increase upon complexation. Upon the addition of glucose as a competitive binding agent, the viscosity decreased significantly while only minor changes were observed upon the addition of others similar chemical substances.

Choi et al. published an analogous boronate-containing gel for insulin delivery based on competitive displacement between glucose and boronic acid binding sites[77].

A similar material, containing both phenylboronic acid and tertiary amine moieties complexed to polyvinyl alcohol, was used as an electrochemical sensor for glucose with a membrane-coated platinum electrode[78]. The polymer complex exhibited enhanced swelling effects proportional to glucose concentration at physiological pH, as glucose displaced the PVA from the boronate groups. Swelling of the cast gel membrane upon glucose addition resulted in enhanced diffusion of ions and a concomitant increase in current.

**15.5 Bio-responsive gels**

The bio-responsiveness of gels can be achieved by different mechanisms that can be classified: 1) change of electrostatic and small-ion osmotic pressure, 2) change of network elasticity by variation of the number of its cross-links and 3) conformational change of polymer chains between cross-links.

The most common examples are given by the glucose-responsive hydrogels investigated for the treatment of diabetes. For instance Hoffman et al.[76] have synthesized a netwoek of poly-glucosyloxy-ethylmethacrylate (PGEMA) interacting with Concavalin A (ConcA) present in the interstitial solution. In presence of competitive free glucose molecules, the number of the PGEMA-ConcA binding sites decreases, inducing gel swelling.

Another type of bio-sensitive gel makes use of natural proteins incorporated into its matrix undergoing conformational transition. An example is given by Mrksich et al.[79] who prepared a gel by covalently linking the Calmodulin (CaM) protein to polymer network by means of PEG molecules. The gel undergoes shrinking when a ligand such as trifluoperazine (TFP) diffuses into the gel and CaM collapses from the dumb-bell shape to a random coil. Since TFP is a calcium-binding ligand, when the calcium is depleted from the interstitial solution the gel swells quasi-reversibly to the initial volume.

Even if several bio-actuated gel systems have been realized[80-81] and many others can be designed, the deficiency of specificity to the target substance may limit their use. To overcome this difficulty, enzyme as well as antigen responsive gels have been synthesized and investigated.

### 15.5.1 Enzyme responsive gels

A glucose-induced gel response can be achieved by immobilizing glucose oxidase together with the catalase protein. Into the gel matrix the first enzyme converts glucose into gluconic acid and hydrogen peroxide in presence of oxygen, while the latter hydrogen peroxide into water and oxygen. In this case, a polymer gel matrix containing ammine groups[82] undergoes swelling as a consequence of the pH lowering generated by gluconic acid.

Another hydrogel system sensible to protease has been synthesized by Ulijn et al.[83]. The hydrogel has aminoacid chains with specific protease-cleavable peptide sites. When the protease is introduced into the gel the anionic peptide is detached and it freely diffuse out the gel leaving the cationic peptide attached to the polymer matrix. The decrease of cross-links of the gel network and the generation of the fixed charged groups on it, leads to a noticeable swelling.

A more complex hydrogel system, conceptually similar to the previous one, has been synthesized to detect elastase[84]. The irreversibility of these kinds of actuation mechanism does not make them useful for repeated release.

### 15.5.2 Antigens and ligands responsive gels

Antigen sensitive gels have been recently synthesized[21,85] by incorporating in the gel network both the antigen as well as the antibody. The antigen-antibody binding increases the cross-links density of the polymer matrix and induces gel de-swelling. When the gel is exposed to a solution with the free antigen that competitively binds itself to the antibody, the number of cross-links decreases and the gel undergoes swelling and it raises its porosity.

More recently, a tumor marker-responsive gel that exhibited volume changes in response to the tumor-specific marker glycoprotein (α-fetoprotein, AFP) was prepared by using lectin and antibody molecules as ligands[86-87].

## 15.6 Biomedical applications

### 15.6.1 Gel actuators

The analogy between the natural muscle and a contractile gel resides just in the outward performance where the actuator element is a material that increases its elastic modulus and shortens its equilibrium rest length to engender force and displacement. Beyond this, the similarity ends. On microscopic scale the mechanism that generates movements and forces in natural muscle is very different from that one of gels.

The biological muscles almost base their unattainable performance on the actin-myosin highly organized system with a maximum stress generation of 300 kPa, contractile strain of about 25% and 50 W/Kg of aerobic power generation that can reach 200 W/Kg in a peak supply.

Even if gels are not able to mimic the muscle, the gel contractility is useful and the nature utilizes it in some cases. For instance, the sea cucumbers, starfish and other echinoderms that embody natural hydrogels made by proteoglycans, elastin, collagen and muscle fibers can quickly switch their elastic properties from soft to hard[88-91]. The elastic change is due to the release of proteins that temporarily bond to the collagen fibers of the matrix.

Even if the gel actuator is not jet a reliable solution as muscle-like engine, laboratory prototypes that attain the muscle performance with a contraction time of 1 second have been realized by using macro-porous NIPAM gel[92-93] and PAN fibers[94-95].

As shown by macro-porous NIPAM actuators, the thermal stimulation is highly reversible but the heat dispersion, when used in the human body, is very slow due to the small thermal gradient that can be esthablished with the surrounding living tissues.

The chemical actuation of PAN fibers gives reproducible force-elongation cycles (figure 3) but the delivering of HCl and NaOH solutions requires a complex system of pumping and piping with a relevant lost of chemical energy due to the unused reagents wetting the surface of the gel fibres.

The electrical activation (figure 4) has engineering problems too. The generation of acid fronts in one gram of acrylic gel needs about 10 mM of cations and a current of about 15 amperes for one minute. If irreversible reactions with production of gas occur at the electrodes, the life of the actuator is very limited. Moreover, degradation of the gel (PAN fibers) due the These problems have received a first solution by the "bi-morph" gel actuators where a thin layer of an ionic polymer gel is enclosed between two thin metal electrodes[98]. This configuration has the advantage to utilize also the electrochemical reaction of the counter-electrode. The ionic fronts of opposite characteristics (e.g., acid and the basic ones) of the electrodes produce the bending of the device due to the gel swelling, on one side, and its de-swelling on the opposite one (see figure 5).

The thin actuator shape leads to fast a response, relatively low currents and energy dissipation. The most promising application for this type of actuator is the realization steerable microcatheter[99] but also Braille display and tactile stimulator are proposed. Recently, the "bi-morph" actuator has been improved by using electrodes made by a network of carbon nano-tubes[100]. It has also been shown that such actuators respond to a mechanical deformation with the generation of an electric potential allowing in principle an inverse sensing function[101].

On the same principle of the "Bi-morph" actuator, an improved layered actuator consisting of multiple thin elements each one composed by two films of gels of opposite chemical characteristic (e.g., a polyacid-polybase gels) has been proposed[102]. The two gel films are sandwiched between two electrodes of conducting polymer containing doping ions that are released (and/or adsorbed) during the electrical inputs (see figure 6). This actuator has the worth of avoiding the gas generation at the electrodes since the electrochemical stimulation is brought by the release of doping ions. Moreover, given that the gel attached on the electrode cannot freely contract in the radial direction, due to the high rigidity of the electrodes, the gel contraction in the axial direction results magnified. The thin layered assembling leads to a fast response as for the bi-morph actuator. The practical limit is constituted by the large number contractile elements, of about 200 per centimeter, needed by a macroscopic actuator.

Given that the chemo-mechanical gel dynamics is mostly diffusive (the electric field is damped by the high concentration of ions), the layers thickness needed to have a contraction of one second is of order of 10 μm for gel with shear and bulk elastic moduli of about 1 Mpa and 5 Mpa respectively, and a friction coefficient of $10^{16}$ Ns/m$^4$. Of the same order of magnitude (10 μm), is the characteristic length of the structure of the natural muscle whose limiting dynamics is given by the diffusion of calcium ions.

Even if an electro-mechano-chemical gel actuator might be shortly realized, the amount of the energy needed to power would constitute a great limitation to its use. The diffusive dynamics of the gel contraction result in a large dissipation into the viscous flow of the gel interstitial fluid.

Arndt and its collaborators[103] have shown that a polyvinylalchol-polyacrylic-acid gel actuator utilizing 175 J/Kg of energy (more than the double of the natural muscle) during its contraction gives an output of 2 W/Kg compared to the 200 W/Kg of the biological analog (figure 7). The efficiency of the gel-like contraction is less than one hundreds of the natural one.

In order to increase the gel actuator efficiency (delivered power) the shortening of its response time becomes very important since a fixed amount of energy is dissipated during each cycle of contraction. Given that the gel contraction scales with the square of its physical length, it is clear that the realization of an efficient gel actuator is based upon the development of a micron-structured system.

In the frame of the present state of art, the mature applications for gel actuators are almost addressed to "low energy" tasks such as valves, light modulators, drug delivery systems, sensing functions and so on, described in the next sections.

### 15.6.2 Gel sensors

The intrinsic energy transduction properties of polyelectrolyte gels and the versatility of functionalized intelligent gels have been widely exploited to design soft sensors for a large variety of measurands.

Sensors for physical, chemical and biochemical measures either as stand-alone devices or fully embedded into close loop systems have been reported.

Possibly the least investigated gel sensors are those intended to detect physical quantities; this fact may originate from the scarce involvement of engineers in the use of soft, wet materials. Besides the above comments polyelectrolyte gel sensors have been described to sense dynamic contact forces, mimicking mechanoelectrical (streaming potentials) properties of human dermis[104]. Other tactile sensors which exploit gel ionization caused by mechanical deformation have also been reported[105-106]. The softness and skin-like mechanical properties of water-swollen gels were claimed to be important whenever gentle object grasp is important, such as in prosthetics and humanoid robotics.

Temperature sensors using intelligent gels have also been described, based on temperature-modulated fluorescence in a polyvinylalcohol/borax hydrogel system containing 2-napthhol[107].

Tunable Bragg[108] reflections in photonic gels have also been exploited to sense different parameters including salt concentration, pressure and humidity[109].

Substantial work has been reported on the use of ionic conduction polymers and polyelectrolyte gels[110].

Work in this area is still intense, trying to overcome problems such as poor stability, hysteresis and slow response.

The use of intelligent gels to implement chemical sensors is essentially related to the detection of the amount of swelling or shrinking of gel based system using different methods in response to the presence of chemical analytes in its aqueous phase. The most investigated gel chemical sensors are those measuring pH[111] where reversible swelling and shrinking is quantified by a piezoresistive element sensitive to gel swelling pressure changes[112]. A $CO_2$ sensor was also developed as a medical indwelling probe based on a pH sensitive hydrogel whose response is governed by the $CO_2$/bicarbonate equilibrium in water[113].

Colloidal crystal hydrogel films has also been developed and use to sense different chemical analytes, based again on Bragg diffraction peaks shifts generated by reversible swelling and shrinking[114].

The use of stimulus-responsive gels as sensor element has found in sensing biochemical analytes its largest interest[115]. A very broad range of biomolecules of medical interest has been investigated, the most relevant being glucose due to the strong need of a sensor to close the loop in insulin delivery systems[116].

Since early studies[117] the use of intelligent gels sensitive to pH changes generated by enzymatic conversion of glucose to gluconic acid has attracted large research interest. Subsequently boronic acid derivatives were used to sense glucose to generate reversible gel expansion and contraction, because of the higher stability of the chemical ligand in comparison to glucose oxydase[118].

Holographic glucose sensors for body fluid measurements[119] and photonic gel sensors[120] for non-invasive monitoring of glucose in tears have been recently reported, both based on boronic acid chemistry.

The literature in this field is broad and the reader may refer to specialized review to get more insight[121]. Another important biochemical analyte whose detection has been accomplished through sensitive polyelectrolyte gels is glutathione, a polypeptide having an important role in several cellular processes whose controlled delivery has therapeutical relevance. A dual responsive and delivery systems has been recently proposed which uses both pH and glutathione sensitive gels to properly tune the release of trapped oligodeoxynucleotides[122].

Antigen sensing through antibody and antigen grafted onto the chains of an intelligent gel has been accomplished through competitive binding resulting in volume changes caused by reversible, non covalent crosslink breaking[21]. Other protein-ligand recognition systems have been disclosed using stimulus-responsive gels and more work is expected in view of the strong interest in biosensor in biotechnology-related areas[123].

### 15.6.3 Gel microfluidic circuits

Sensors, channels, valves and pumps are the basic components of microfluidic circuits which are needed for the development of lab-on-a chip systems for biochemical analysis[124-125], genomics and proteonomics[126] and cell studies[127].

In the early 1980's the exploitation of silicon technology and the use of silicon as a mechanical material[128] has provided powerful tools to the fabrication of MEMS and BioMEMS for miniaturised analytical systems on a chip[129-130]. The manipulation of fluid at the microscale to implement fully operative separation and analytical system has been since then very intensively studied[131]. The main drive for lab-on-chip development still is the need for cheap, reliable, simple and even disposable analytical systems at the point-of-care.

Despite tremendous R&D efforts, however, still several technical obstacles impede the full exploitation of BioMEMS devices. Possibly, the major difficulties rely on the fact that the superior mechanical properties of silicon and the powerful microfabrication technique nowadays available are largely vanified by the absence or inefficiency of silicon transduction properties. These properties are absolutely necessary to realize active components such as pumps, valves and sensors.

The advent of hybrid technologies which rely on integration of silicon with other solid state materials endowed with piezoelectric, photoemissive and other transduction properties is complex and expensive.

Stimulus-responsive gels have been proposed and intensively studied as possible alternative materials that can be easily microfabricated and possess all the needed properties of sensing, actuation and even self-regulation[132].

Besides the easy patterning and microfabrication techniques which can be adopted to build gel active microcircuits, the main advantages of stimulus-responsive gels in this field of application is the possibility to use chemicals dissolved in the liquid phase as stimuli to trigger functions without the need of external action and control[133] or to use convenient photoirradiation with time and space selective control[134].

Thermally[135] and pH[136] activated gel valves for microfluidic chips have been developed as much as micromixers and micropumps[137-138], adjustable focus microlenses integrated into gel microfluidic systems for optical sensing[139-141].

3D patterning and direct writing of intelligent gels microstructures have also been reported[142-143] opening up further avenues toward the achievement of the difficult goal of fully-integrated 3D active gels microfluidic channels.

### 15.6.4 Gel drug delivery systems

The "drug delivery system" is a very wide concept that embraces all means that can lead to a release of drugs to the desired target site, at the right time and in the right dose. This has the advantage to almost avoid side effects, to bring to an efficient use of drugs and possibly to substitute medical interventions by pharmaceutical ones.

Most of the affirmed drug delivery systems are solid polymers to sustain a prolonged release of active substance since hydrogels would release the small molecules very quickly for most of the applications. Recently, the use of drugs made of larger molecules such as proteins and peptides, that are effective at a very low dose, has brought the attention on gels as tools for efficient drug delivery. The nowadays applications use hydrogels for protection of protein drugs (e.g., by the digestive system), for the adhesion of patches releasing drugs and for the protection of nanoparticles of drugs by the immune system[144].

Moreover, the use of the electrical activation to enhance drug outflow by patches or internal gel systems, has been proposed by many authors[145]. Drug eluting stents have also been realized by using thermal-sensitive gels[146]. A more sophisticated electrically controlled system made by gels and conducting polymers for the release of a drug has been realized by Brahim et co-workers[147](figure8). Moreover, the ability of gels (described in the previous sections) to react to the presence of glucose leads to the design of a variety of insulin delivery systems at a controlled rate[76-78].

Finally, it is worth mentioning the great interest in smart gels for encapsulating micro-particles of drug that are absorbed at the target biological site or organ[148-155].

### 15.6.5 Gels for chromatographic and membrane separation

Since the early observations of large changes in solute permeability of biological[156] and synthetic[157] polyelectrolyte membranes upon changes in their ionization state, substantial research activity has been devoted to better comprehend and potentially exploit these properties for solute separation[158]. Different stimuli have been used to alter and control polyelectrolyte gels membranes, the most common being pH or ionic strength [159], electric field[23] and temperature[160] changes. The field has been recently reviewed[161] with particular attention to separation by chromatographic techniques through stimulus-responsive gels.

Most of the work related to solute separation by intelligent polyelectrolyte gels has been addressed to chromatographic separation through ionic or hydrophobic interactions[162-164], temperature-responsive stationary phase for size exclusion[165], and affinity-based separation[123].

Temperature control of gels mostly affects the hydrophilic-hydrophobic balance of the network and average pore size. On the other hand, pH, salts, and electric fields mostly affect ionic interactions and Donnan partition.

Conjugating temperature-responsive gels with ligands offers a very interesting way of modulating non-covalent interactions between stationary and mobile phases without recurring to changes in buffer composition or to other cumbersome techniques. A large variety of biomolecules have been selectively separated through affinity-modulation by smart gels[166] potentially providing cost effective and simple separation of biologically active compounds.

*15.6.6 Gel tissue analogs*

The ubiquity of gels in natural systems and biological tissues has been a great source of interest for these materials. The bi-phasic architecture of gels that displays both a solid-like rheology (undergoing to morphological changes) and hosts electro-convective and chemo-diffusional processes, typical of fluids, elicits a series of complex biological functionalities.

The comprehension of gel systems has lead to a better understanding of many biological phenomena[167-169]. Obviously, the knowledge about gels comes through an approximated model that is a synthetic and incomplete representation of real systems.

Between the theoretical "biphasic" material and the real system of a biological tissue there are two main differences. The "bi-phasic" model, shown in the preceding sections, was developed by Biot for the study of impregnated rooks in the geological field. Taken as it is, in principle it does not make any differences with gels.

If the low frequency model gets a satisfying experimental validation and gives interesting insights about biological phenomena such as the streaming potentials in bone[182], the sensing ability of skin[167], the compressional behavior of cartilage, and so on; the response of gels to ultrasounds deviates by that of the standard poroelastic theory[33]. As shown recently, the response of gels to ultrasounds is sensibly influenced the presence of bounded water[170].

Moreover, even if a sophisticated bi-phasic model can describe the gel behavior, the tissue usually host an amount of cells that own a membrane with a jelly internal structure whose poroelastic properties can differ much by the extra-cellular matrix.

Therefore the model of a soft biological tissue must introduce a syncytium of dispersed cells into the gel matrix[171]. In this case, is possible to define an overall poroelasticity of the composed mean that depends by the extra-cellular matrix characteristics, by the cell membrane permeability, by the cell internal poro-elasticity and its fractional volume. Since all these variables depend by the physiological state of the cells in principle the ultrasound tissue response can be correlated to its health state. The possible applications in this field are the development of non-invasive methodologies for health tissue evaluation such as elastography for cyrrosis[172] and bi-phasic imaging[173].

An accurate model of gel response allows also the realization of phantoms by which to replicate and to investigate biological phenomena in laboratory in a repetitive manner. The main applications in this field are the ultrasound thermal therapy for tumors[174-175] and the focused ultrasound surgery[176]. Rovai and its collaborators[177] have studied the effect of cavitation ultrasound micro-bubbles on thrombi in an experimental cell by using a gel tissue phantom. Also pulmonary comets[178] can be replicated by means of a gel phantom mimicking the lung structure leading to quantitative analysis of the phenomenon and its correlation with the water present in the tissue[179] and its origin.

## 15.6.7 Gels for cell culture and tissue engineering

"In vitro" cell and tissue culture have experienced tremendous advances in the last 20 years mostly due to the discovery and exploitation of new techniques in cell pattering and printing[180] and tissue engineering[181]. Cell encapsulation, cell sheets adhesion modulation and 2D and 3D polymer scaffolding for cell support all have benefited of the use of hydrogels[182] and responsive gels[183]. Temperature and biomolecules responsive hydrogels are currently used to endow extracellular artificial cell matrix with specific functional properties that consent controlled surface modification or release of bioactive compounds to interact with the cellular component. Temperature sensitive gels and their application to cellular engineering have been pioneered by T. Okano[184-172]. Dramatic changes in wettability of PIPAm grafted onto cell culture substrates[173] after temperature variation from 32 °C (PIPAm LCST) and 20 °C have been exploited to detach cell culture from their substrate without using digestive enzymes or chelating agents. Cells adhere and proliferate onto PIPAm treated culture dishes at 37°C when the surface is hydrophobic and they detach when temperature is lowered to 20°C when the surface becomes hydrophilic[187].

Based on this technique an alternative approach to 3D scaffolding for tissue engineering has been proposed with the given name of "cell sheet engineering"[188] which promises to broaden the capabilities and applicability of tissue engineering in the clinical setting[189-190].

The process of biomineralization has also been shown to be affected by temperature sensitive hydrogels when PIPAm was grafted onto poly(L-lactic) acid and Bioglass[191]. Apatite was found to form onto the substrate when the material was kept at the temperature of 37°C, above LCST of PIPAm while no precipitation occurred below 32°C. Poly(L-Lactic) acid and Bioglass substrates were modified by grafting chitosan through plasma treatment and pH chitosan responsively was exploited to trigger apatite deposition[192].

**Appendix**

*A.1 Network readjustment kinetics*

When the equilibrium conditions change, the gel starts to readjust itself to a new stationary state.

Given the multi-variable nature of the gel free energy, the readjustment kinetics may concern a large number of thermodynamic variables such as temperature, volume, pressure, length, mechanical stress, concentration of chemical species of the interstitial solution and electrical potential.

When such a system changes its state, forces and related fluxes of all set of extensive-intensive variables appear. Generally speaking, the kinetics are coupled each other and the overall system of equations is not manageable.

In certain cases, these kinetics have very different characteristic times so that the related thermodynamical co-ordinates can be considered quasi-constant or at equilibrium during the process allowing simplified descriptive models[32].

For instance, if the diffusion of the chemical species into the gel network is much slower than its mechanical readjustment, the gel volume can be considered at equilibrium with the local chemical conditions.

In other cases, under controlled laboratory condition, some of the gel variables can be held constant allowing the investigation of one type of readjustment kinetic at a time. This is the case of free swelling experiments where it is possible to observe the mechanical readjustment of the gel network while the gel temperature and its chemical environment are held constant.

By means of laboratory-controlled experiments it is possible to determine the chemical and elastic constants of the gel present into the theoretical model. Moreover, when the gel deformations occur on a scale much larger than the molecular one, the continuum approach can be assumed.

Even with some limitations[33], continuum poroelastic models have shown to satisfactory describe the readjustment of a gel system[193-194].

In the low frequency limit, the Biot's continuum poroelastic model [195] is a simple and elegant theory that also describes the diffusion kinetics of gels.

In gel with diluted polymer (solid) content and with incompressible fluid and solid constituents, the mechanical readjustment of the gel network is satisfactorily described by the stationary solvent approximation. In this case, the Biot model disembogues in the THB[196] frictional equation

$$f\, \partial U_i / \partial t = \partial \sigma_{ij} / \partial x_j \qquad (A.1)$$

where $U_i$ is the displacement vector of a gel element, f is the gel friction coefficient (the inverse of the gel hydraulic permeability[197]), $\sigma_{ij}$ is the gel stress tensor that in the linear approximation reads:

$$\sigma_{ij} = k\, \varepsilon_{\alpha\alpha}\, \delta_{ij} + 2\mu(\varepsilon_{ij} - \varepsilon_{\alpha\alpha}\, \delta_{ij}/3) + \alpha\, \delta_{ij} \qquad (A.2)$$

where k and $\mu$ are respectively the bulk and the shear elastic moduli of the gel, $\alpha$ is the chemically or thermally induced stress (at zero strain) for isotropic materials, $\delta_{ij}$ is the Kroneker delta and $\varepsilon_{\alpha\alpha}$ is the gel dilatation given by the trace of the strain matrix $\varepsilon_{ij}$

$$\varepsilon_{ij} = (\partial U_i / \partial x_j + \partial U_j / \partial x_i) / 2 \qquad (A.3)$$

Generally speaking, the material parameters in Equation (A.1) and Equation (A.2) are functions of the physical variables of the material (e.g. temperature (see figure 9 and figure 10), strain, etc.) as well as of the chemical ones (e.g., pH, ionic strength, type of solvent and so on).

Because the dependence of the material parameters on the mechanical deformation is usually weak, $\mu$, k, and f can be assumed constants if the gel state is far away from a volume phase transition.

By introducing Equation (A.2) in Equation (A.1), by taking the divergence of both members and inserting the incompressibility of solid and liquid constituents, the equation of motion finally reads [198]:

$$f \, \partial \varepsilon_{\alpha\alpha} / \partial t = (k+4\mu/3) \, \partial^2 \varepsilon_{\alpha\alpha} / \partial x_i \, \partial x_i \qquad (A.4)$$

Actually, the poro-elastic model is a semi-empirical model, where the link between the material parameters and the physico-chemical structure of the gel is not explicit.

This fact allows the model to hold for a large typology of gel systems, where the dependence of the material parameters by the environmental conditions is usually experimentally measured.

As a first example, we consider the free-swelling experiments, a well-known technique that allows characterizing the poro-elastic constants of a gel system. In such tests an initially compressed gel network is let to freely expand in a bath held under constant physico-chemical conditions.

In the case of gel shapes with particular symmetries such as in form of a sphere, thin cylinder or a thin planar film, equation (A.4) can be simplified and it leads to analytical solutions for the time evolution of the local gel strain[199-200]. Here we report the solution concerning the case of gel in form of thin planar film or disk. When the gel sample has the shape of a thin quasi-planar layer, assuming the z-axis perpendicular to the gel layer plane, Equation (A.2, A.4) can be simplified to read [201].

$$\partial \varepsilon_{xx} / \partial t = D \, \partial^2 \varepsilon_{xx} / \partial z^2 \tag{A.5a}$$

$$\partial \varepsilon_{yy} / \partial t = D \, \partial^2 \varepsilon_{yy} / \partial z^2 \tag{A.5b}$$

$$\partial \varepsilon_{\alpha\alpha} / \partial t = D_b \, \partial^2 \varepsilon_{\alpha\alpha} / \partial z^2 \tag{A.5c}$$

Where $D = \mu / f$ and $D_b = (k+4\mu/3) / f$.

The spatio-temporal solutions for the strains $\varepsilon_{xx}$ (figure 11) and $\varepsilon_{zz}$ (figure 12) are[199]:

$$\varepsilon_{xx} = \frac{4\varepsilon_0}{\pi} \sum_{n=1}^{\infty} \left(\frac{(-1)^n}{2n+1}\right) \exp\left[\frac{(2n+1)^2 t}{\tau}\right] \cos\left(\frac{(2n+1)z\pi}{a}\right) \tag{A.6a}$$

$$\varepsilon_{zz} = \varepsilon_{\alpha\alpha} - 2\varepsilon_{xx} = \frac{4\varepsilon_0}{\pi} \sum_{n=1}^{\infty} \left(\frac{(-1)^n}{2n+1}\right) \cos\left(\frac{(2n+1)z\pi}{a}\right) \left\{3 \exp\left[\frac{(2n+1)^2 t}{\tau_b}\right] - 2 \exp\left[\frac{(2n+1)^2 t}{\tau}\right]\right\} \tag{A.6b}$$

Where $\tau = a^2 / \pi^2 D$ and $\tau_b = a^2 / \pi^2 D_b$ are the characteristic time constants for the "shear" and "bulk" gel readjustment where $\tau_b < \tau$, since $D_b > D$; $\varepsilon_0$ is the initial uniform strain of the gel sample with respect the final one (at t = ∞) assumed as reference ($\varepsilon_\infty = 0$) and "a" is the gel layer thickness at t = ∞.

From Equations (A.6a,b) the thickness $a_{(t)}$ (figure 13) and the length $L_{(t)}$ (figure 14) of the gel are obtained as a function of time, respectively, to read 199.

$$a_{(t)} = a\{1 + 2\int_0^{a/2} \varepsilon_{zz}\} = a\{1 + \frac{8\varepsilon_0}{\pi^2} \sum_{n=1}^{\infty} (\frac{(-1)^n}{2n+1})\{3\exp[\frac{(2n+1)^2 t}{\tau_b}] - 2\exp[\frac{(2n+1)^2 t}{\tau}]\}\} \quad (A.7a)$$

$$L_{(t)} = L_\infty \{1 + \varepsilon_{xx}(Z=0)\} = L_\infty \{1 + \frac{4\varepsilon_0}{\pi} \sum_{n=1}^{\infty} (\frac{(-1)^n}{2n+1})\exp[\frac{(2n+1)^2 t}{\tau}]\} \quad (A.7b)$$

For $t > \tau/9$ in Equation (A.7a) the slower exponential relaxation prevails so that the gel length reads:

$$L_{(t)} \cong L_\infty \{1 - \frac{4\varepsilon_0}{\pi}\exp[\frac{t}{\tau}]\} = L_\infty \{1 - \frac{4\varepsilon_0}{\pi}\exp[\frac{\pi^2 D}{a^2} t]\} \qquad t > 9\tau \qquad (A.8)$$

By fitting the exponential length relaxation of the gel it possible to obtain the characteristic time $\tau$ and the gel diffusion coefficient D.

## A.2 Electrodiffusion-reaction kinetics

In the case when the concentrations of the chemical species may change in the gel bath, the presence of gradients will generate chemical currents in the interstitial solution.

Given that the poroelastic parameters $\mu$, k, f, $\alpha$ depend from the chemical concentrations $C_i$ of the interstitial solution (see for instance figure 15), their redistribution may induce mechanical swelling or de-swelling of the gel system.

In this case the mechanical equations (A.1-3) are coupled to the Nernst-Plank electro-convective equations for the motion of each charged chemical species (together with the electrical charge conservation and the Gauss equation for the definition of the electric field)[202].

The electroconvective kinetics introduce a mathematical complexity that usually can be circumvented when they are very fast with respect to the other ones so that the electrical equilibrium (null electrical currents) or the stationary electrical conditions can be imposed.

Assuming that the convective and migration currents can be disregarded, the full system of equations reads:

$$f_{(Ci)} \, \partial U_i / \partial t = \partial \sigma_{ij} / \partial x_j \tag{A.9}$$

$$\sigma_{ij} = k_{(Ci)} \, \varepsilon_{\alpha\alpha} \, \delta_{ij} + 2\mu_{(Ci)} \cdot (\varepsilon_{ij} - \varepsilon_{\alpha\alpha} \, \delta_{ij} /3) + \alpha_{(Ci)} \, \delta_{ij} \tag{A.10}$$

$$\varepsilon_{ij} = (\partial U_i / \partial x_j + \partial U_j / \partial x_i) / 2 \tag{A.11}$$

$$\partial C_i / \partial t = -D_{(C_i)} \partial^2 C_i / \partial x_j \partial x_j + \partial C_{i\,chem} / \partial t \tag{A.12}$$

where $C_{i\,chem}$ is the concentration in moles per liters of the i-th specie that has reacted.

In the above equations we have considered that the temperature is constant through all the process.

Actually, temperature can vary as a consequence of chemical reactions as well as of external inputs as it happens in thermally activated gel systems. In this case the equation of thermal diffusion must be added to the system of equations.

The resulting overall system of equations is complex and usually cannot be treated analytically.

Nevertheless, some insight can come by investigating the simplest case of diffusion of hydrogen ions in a gel matrix with ionizable groups (RAH) undergoing dissociation

$$RAH \Leftrightarrow RA^- + H^+ \tag{A.13}$$

In this case, the system of equations reads

$$f_{(H^+)} \partial U_i / \partial t = \partial \sigma_{ij} / \partial x_j \tag{A.14}$$

$$\sigma_{ij} = k_{(H^+)} \varepsilon_{\alpha\alpha} \delta_{ij} + 2\mu_{(H^+)} \cdot (\varepsilon_{ij} - \varepsilon_{\alpha\alpha} \delta_{ij}/3) + \alpha_{(H^+)} \delta_{ij} \tag{A.15}$$

$$\varepsilon_{ij} = (\partial U_i / \partial x_j + \partial U_j / \partial x_i)/2 \tag{A.16}$$

$$\partial [H^+]/\partial t = -D_{(H^+)} \partial^2 [H^+]/\partial x_j \partial x_j - \partial [RAH]/\partial t \tag{A.17}$$

Usually, the chemical reaction is faster than the gel mechanical readjustment as well as the chemical diffusion so that the reaction equilibrium condition can be applied to read

$$K_A^* = [RA^-][H^+]/[RAH] \tag{A.18}$$

Where $K_A^*$ is not a simple reaction constant since in the gel matrix the ionization of each functional groups is influenced by the state of the groups in its neighborhoods [203].

Therefore, KA* depends on the idrogen ion concentration and it is related to the free acid equilibrium constant $K_A = [A^-][H^+]/[AH]$ by the relation

$$K_A^* = [H^+] + K_A \qquad (A.19)$$

Moreover, given that the variation of the chemical stress $\alpha_{(H^+)}$ as a function of the proton concentration is much bigger than those ones of $f_{(H^+)}$, $k_{(H^+)}$ and $\mu_{(H^+)}$, at the zero order of approximation we can consider the latter ones as constants to end with the motion equations

$$f\, \partial U_i / \partial t = \partial \sigma_{ij} / \partial x_j \qquad (A.20)$$

$$\sigma_{ij} = k\, \varepsilon_{\alpha\alpha}\, \delta_{ij} + 2\mu\, (\varepsilon_{ij} - \varepsilon_{\alpha\alpha}\, \delta_{ij}/3) + \alpha_{(H^+)}\, \delta_{ij} \qquad (A.21)$$

$$\varepsilon_{ij} = (\partial U_i / \partial x_j + \partial U_j / \partial x_i)/2 \qquad (A.22)$$

$$\partial [H^+] / \partial t = -D_{eff}\, \partial^2 [H^+] / \partial x_j \partial x_j \qquad (A.23)$$

$$D_{eff} = D_{(H^+)} / (1+ [RA^-] K_A / ([H^+] + K_A)^2) \qquad (A.24)$$

$$[RA^-]_{(x,t)} = [RA^-]_0 \, \varepsilon_{\alpha\alpha \, (x,t)} \qquad (A.25)$$

where $[RA^-]_0$ is the ionizable groups concentration at the initial gel volume.

From the motion equations (A.20-25) we can see that the chemical problem is coupled to the mechanical one through the gel dilatation evolution $\varepsilon_{\alpha\alpha \, (x,t)}$ and that the mechanical dynamics is coupled to the chemical one through the chemical stress $\alpha_{(H^+(x,t))}$.

If we start from a basic condition (swollen polyacid gel with most of the ionizable groups dissociated), as far as $[H^+] \ll K_A \sim 10^{-4 \div 5}$ we can use the approximation[204]

$$D_{eff} \approx D_{(H^+)} / (1+ [RA^-] / K_A) \qquad (A.26)$$

From the above formula when the ionizable group density $[RA^-]$ on the gel network is much larger than the acid dissociation constant $K_A \sim 10^{-4 \div 5}$, it follows that the effective diffusion coefficient may result much smaller than the free proton diffusion one. Initially, in the acidification process (inflow of $H^+$) there is an excess of binding sites available in the gel so that almost all the incoming hydrogen ions are immediately bound to these sites and cannot thus freely diffuse through the gel.

Since the mechanical readjustment, as well as the chemical process, follow diffusive kinetics, their characteristic times, that scale by the square of the same characteristic length of the system, depend on the elastic constant $k_{(H^+)}$, $\mu_{(H^+)}$, on the friction coefficient $f_{(H^+)}$ of the gel network, on the hydrogen diffusion coefficient $D_{(H^+)}$ and on the ratio between the ionizable group density $[RA^-]$ of the gel network and the acid constant $K_A$, respectively. Therefore, in a soft weakly charged gel ($[RA^-] / K_A \ll 1$) with a highly viscous solvent-network interaction it may result

$$D_{eff} \approx D_{(H^+)} \gg D_b = (k+4\mu/3) / f \qquad (A.27)$$

In this case the network motion is much slower that the chemical kinetics so that they decouple themselves: the gel volume change can be described to happen as a consequence of a step change of the chemical conditions to the final ones.

The chemo-mechanical decoupling can also happens in a strong elastic gel ($\mu$ and $k$ very high) with heavily charged network ($[RA^-] / K_A \gg 1$) where it may result

$$D_{eff} \approx D_{(H^+)} K_A / [RA^-] \ll D = \mu / f < D_b \qquad (A.28)$$

In this case the gel kinetics is simplified since its mechanical state is always at equilibrium with the local chemical conditions.

The proton diffusion is a simple example of chemo-mechano-electrical kinetics in gels. More recently, the development of bio-responsive hydrogel for biomedical application has brought to the synthesis of many complex gel systems where the diffusion of a specific chemical specie or biological compound elicit a process leading to the gel volume and poro-elasticity change. Even much more complex, these kinetics follow the scheme of a diffusion of a compound in the gel matrix coupled to a reaction that leads to the readjustment of the network as well as of the ionic concentrations in the interstitial solution. This is the case of enzyme-loaded gels that in presence of the target substance may change the pH of the interstitial solution, or the case of gels using the antigen-antibody binding reaction to change the degree of cross-linking of its network.

## A.3 A non-equilibrium thermodynamics view of electromechanical phenomena

When the gel has a charged polymer matrix, or ions are dissolved in the interstitial solution, in addition to the polymer network motion we have fluxes of ionic charges.

In this case, the hydraulic flux governed by Darcy law is coupled to the electric one (figure 16) that at first order can be described by the Onsager relations that, in the interstitial stationary fluid approximation, read

$$J_i^{netw} = \partial(U_i - u_i^{fluid})/\partial t \cong \partial U_i/\partial t = K_{11}\,\partial P/\partial x_i + K_{12}\,\partial \Psi/\partial x_i \qquad (A.29)$$

$$J_i^{charge} = K_{12}\,\partial P/\partial x_i + K_{22}\,\partial \Psi/\partial x_i \qquad (A.30)$$

where $K_{ij}$ are the electro-osmotic Onsager coefficients [205], P is the hydraulic gel pressure, $\Phi$ is the electric potential, $J_i^{netw}$ is the current of the polymer with respect to the fluid, and $J_i^{charge}$ is the ionic charge current of the i-th specie.

The introduction of the electric field increases the number of variables as well as the number of equations. As already noted in the preceding paragraph, Gauss equation for the electric field as well charge conservation law must be introduced in order to know how the electric potential builds itself up when the electric charges move.

Even if no external electric field is applied, there exists the one generated by the motion of the charges due to the hydraulic pressure gradients in the coupled equations (A.29-30).

The introduction of Gauss equation makes hardier to have a tractable solution of the electro-mechanical problem. Nevertheless, in gels when the ionic conductivity is so high that the electrical equilibrium is much faster than the network diffusive readjustment we can impose that the electric charges are always at the stationary state to read

$$J_i^{charge} = K_{21}\, \partial P/\partial x_i + K_{22}\, \partial \Psi/\partial x_i = 0 \qquad (A.31)$$

This simplifies very much the problem leading to the expression for the electric potential that reads

$$\partial \Psi/\partial x_i = -(K_{21}/K_{22})\, \partial P/\partial x_i \qquad (A.32)$$

from which we obtain,

$$\partial U_i/\partial t = \{K_{11} - K_{12}(K_{21}/K_{22})\}\, \partial P/\partial x_i \qquad (A.33)$$

Moreover, even if the solvent is stationary, in the low frequency limit a non-null force acting on the polymer network exists, which is at equilibrium with the hydraulic pressure gradient to read

$$\partial P/\partial x_i \cong \partial \sigma_{ij}/\partial x_j \qquad (A.34)$$

so that the friction coefficient by equation (A.33) reads

$$f = (K_{11} - K_{12} K_{21} / K_{22})^{-1} \qquad (A.35)$$

It is straightforward to see that f reduces to the inverse of hydraulic permeability $K_{11}$ when no charges are present into the gel system and $K_{12}$ = $K_{21}$= 0.

This electro-mechanical coupling in gels is responsible both for the generation of electrical potentials in presence of gel readjustment induced both by mechanical inputs such as compression (figure 17), deformation and by physico-chemical inputs such as temperature, ph, salt concentration, solvent affinity and so on. Inversely, it is responsible for the gel mechanical response to the electro-chemical currents and potentials.

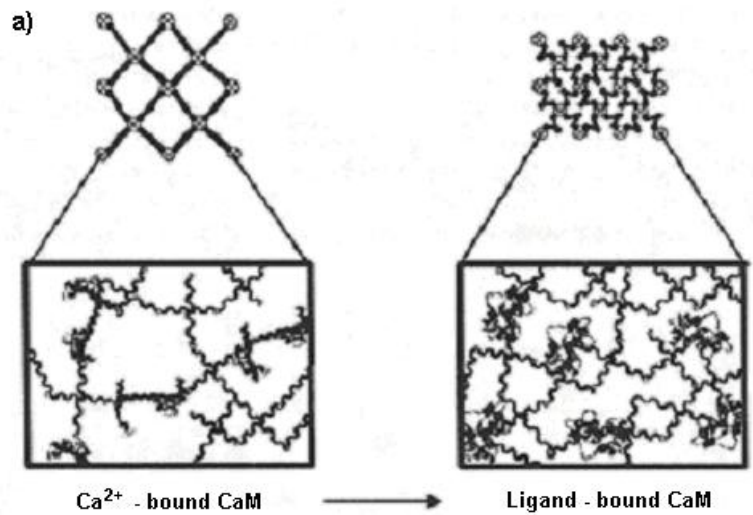

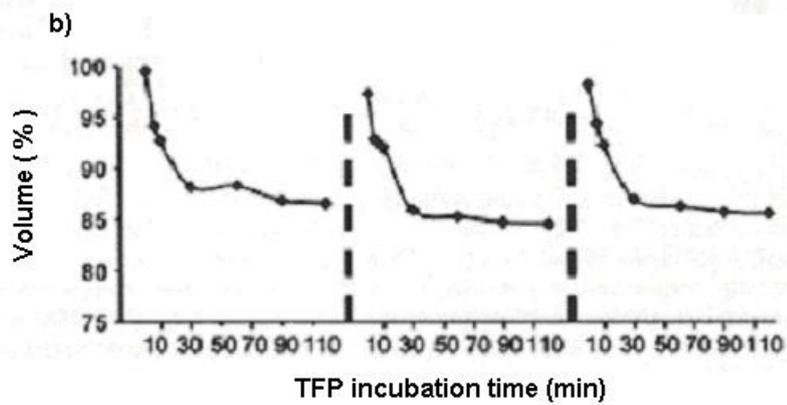

*Figure 1* (a) The extended (left) and collapsed (right) states of CaM in the hydrogel network; (b) The volume of the hydrogel exposed to TPF as a function of time and after washed repeatedly in a calcium containig bath to restore the extended CaM configuration (Reprinted with the permission from Z. M. Yang et al.. Copyright (2006) Am. Chem. Soc.).

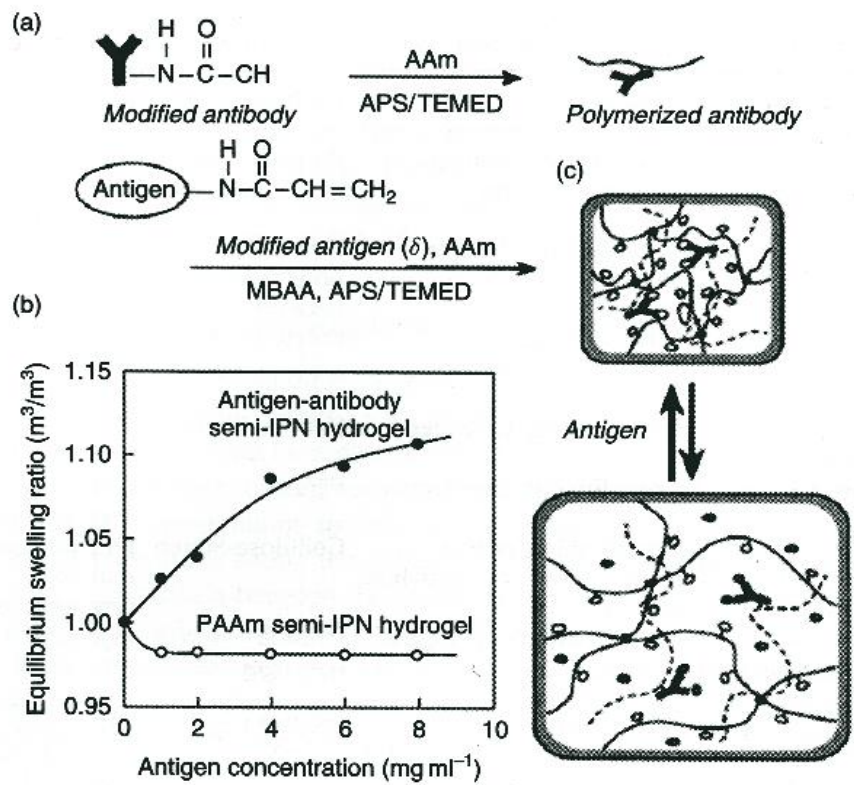

*Figure 2* Antigen-responsive hydrogel. (a) Synthesis of the antigen-bounded network. (b) Hydrogel swelling as a function of the free antigen concentration. (c) Mechanism of the free antigen competitive binding. (Reprinted with the permission from T. Miyata et al.. Copyright (1999) Nature Publishing Group)
70

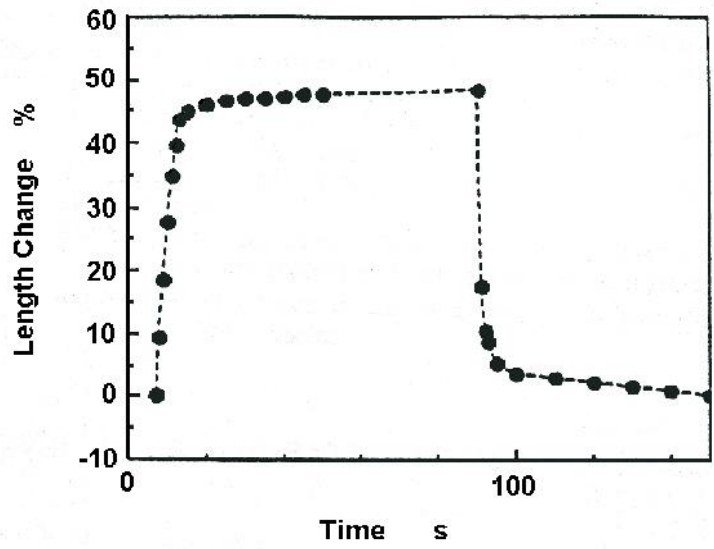

*Figure 3* The isotonic fractional length change of PAN gel fibers as a function of time when the external bath pH is suddenly changed from 1 to 13.

electrochemical potentials is actually observed[96-97].

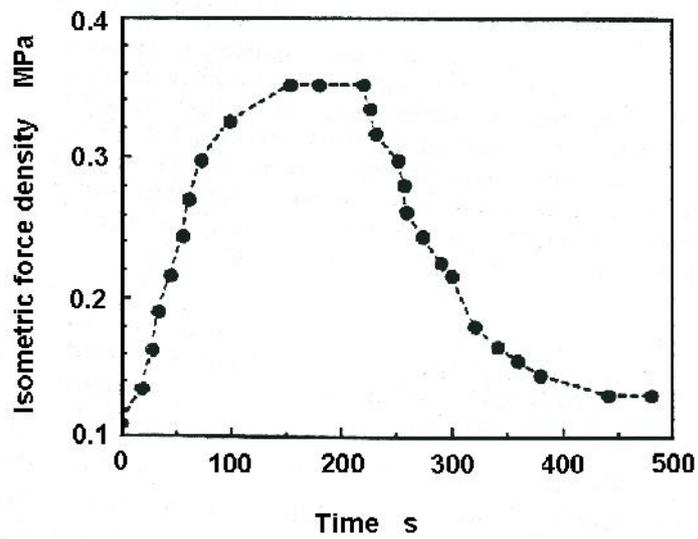

*Figure 4* The isometric force density generated by the PAN gel fibers as a function of time when the carbon fibers electrodes are excited by steps of electric potential between +10 and -10 volts.

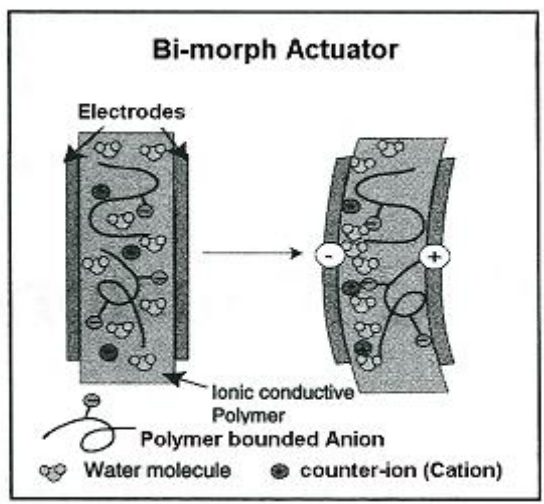

*Figure 5* *Schematic view of an ionic polymer-metal composite (IPMC) Bi-morph gel actuator.*

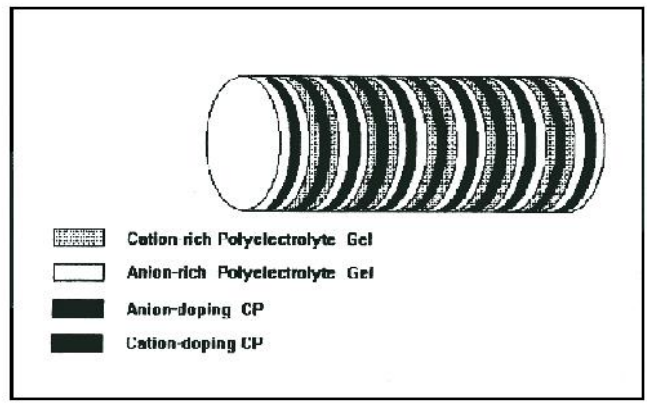

*Figure 6* Schematic view of a layered electrically driven gel actuator.

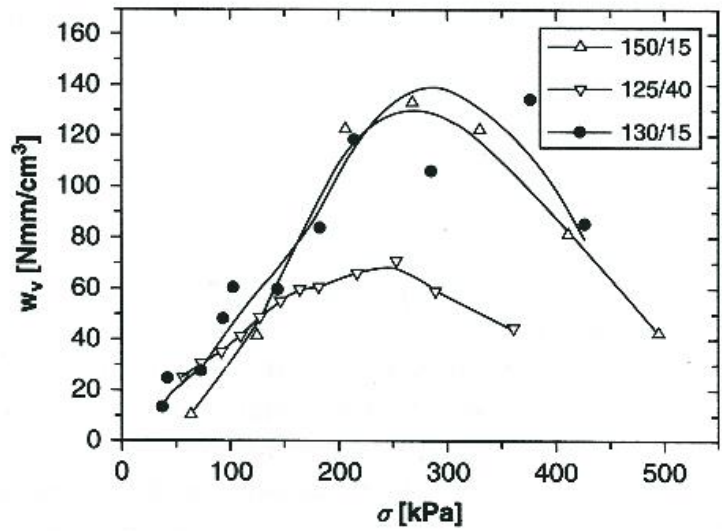

*Figure 7* Work energy generated by PVA-PAA hydrogel under different temperature and duration of cross-linking. (Reprinted with the permission from Ref [103]. Copyright (1999) Wiley-VCH Verlag GmbH).

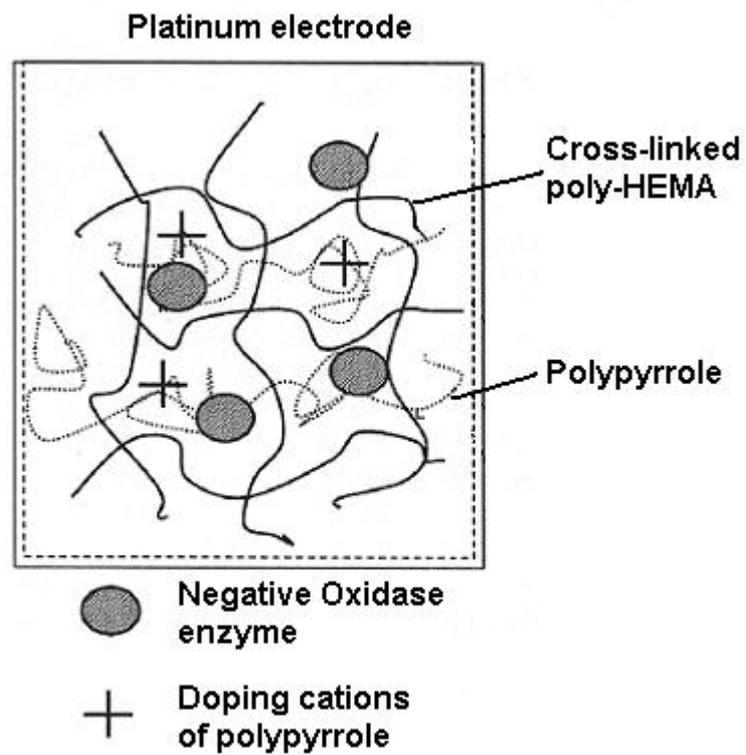

*Figure 8* Cross-linked pHEMA and interpenetrating polypirrole with entrapped GOx coating a platinum electrode for amperometric glucose sensing.

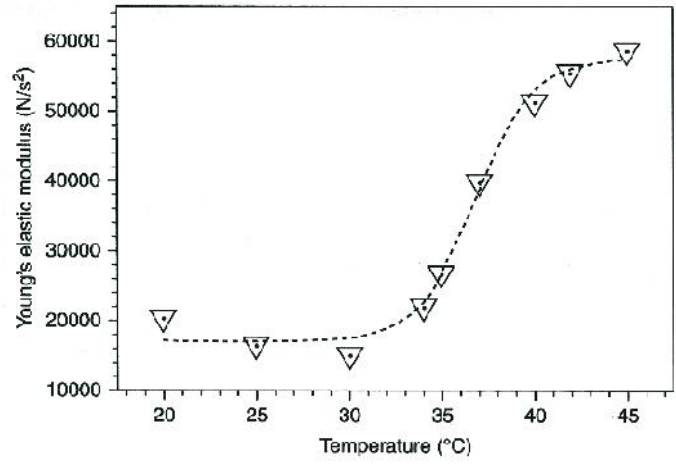

***Figure 9*** *The Young's elastic modulus $E = 9\mu k / (3k+\mu)$ of the PVME macroporous gel as a function of the temperature.*

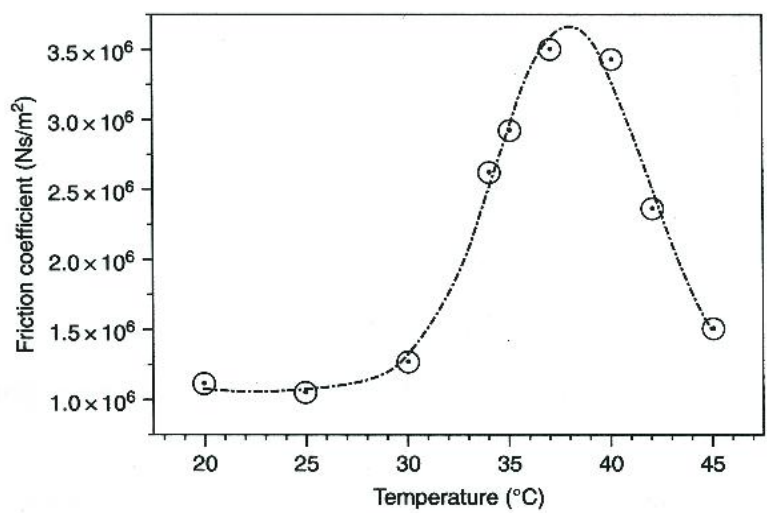

*Figure 10* *The fluid-matrix friction coefficient "f" of the PVME macroporous gel as a function of the temperature*

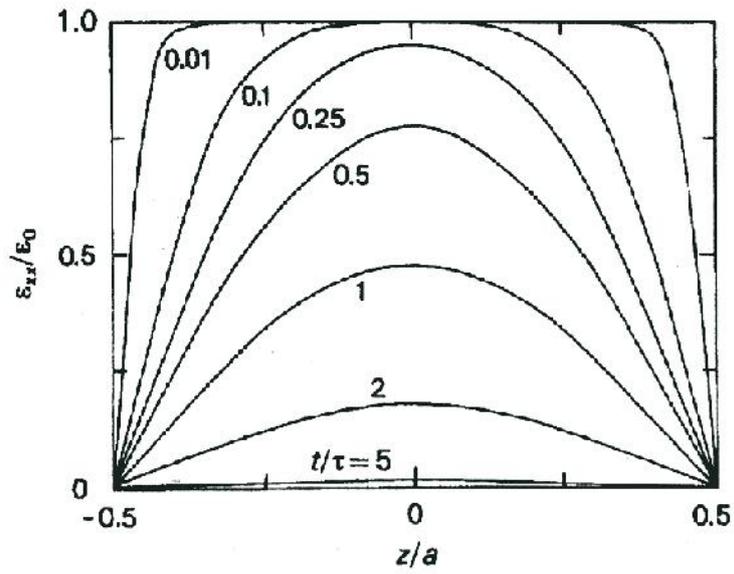

*Figure 11* Normalized strain $\varepsilon_{xx}/\varepsilon_0$ in a partially dried hydrogel strip undergoing free swelling as a function of the reduced variable $z/a$ at various scaled times $t/\tau$ given by equation 6a.

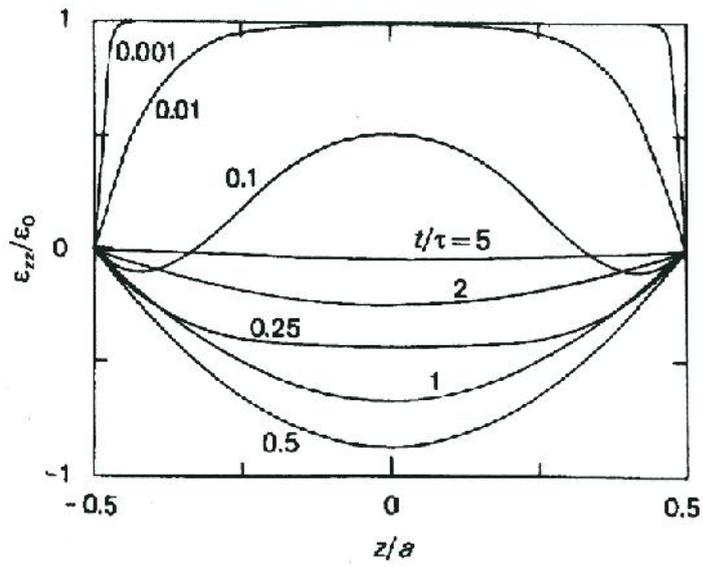

**Figure 12** Normalized dilatation $\varepsilon_{zz}/\varepsilon_0$ in a partially dried hydrogel strip undergoing free swelling as a function of the reduced variable $z/a$ at various scaled times $t/\tau$ given by equation 6b.

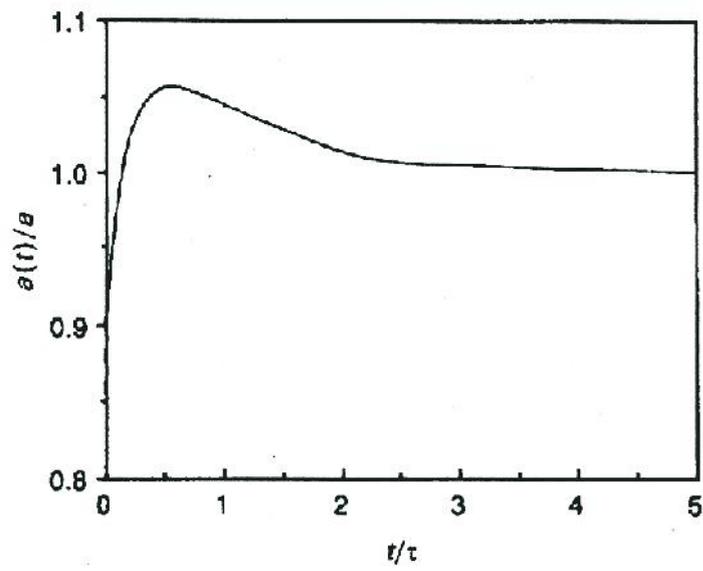

*Figure 13* Normalized thickness of a thin hydrogel strip undergoing free swelling as a function of the reduced time t/τ given by equation 7a.

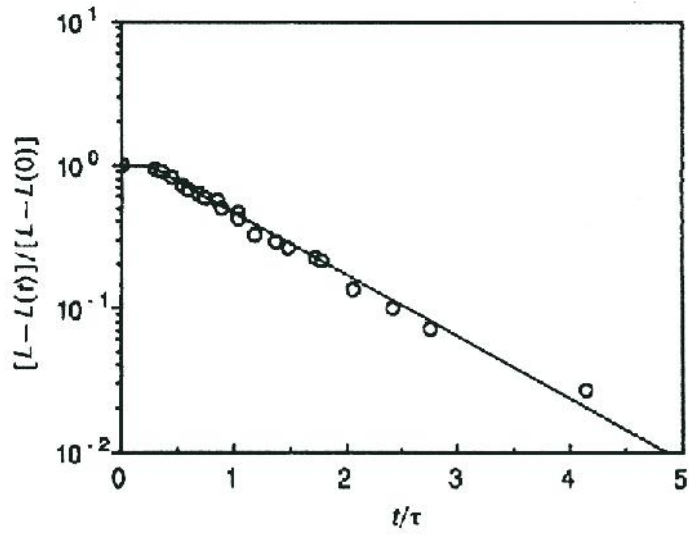

*Figure 14* Normalized length variation of a thin hydrogel strip undergoing free swelling as a function of the reduced time $t/\tau$ given by equation 8.

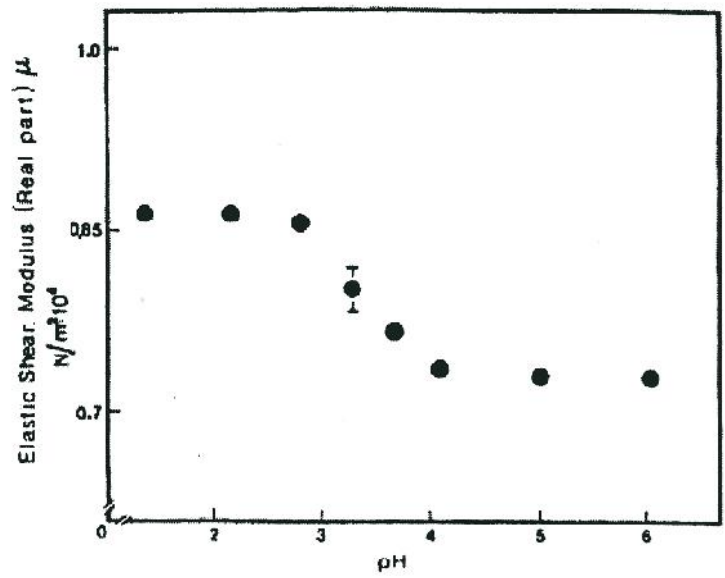

*Figure 15* *The shear elastic modulus of polyvinylalchol-polyacrylic-acid gel as a function of the external bath pH.*

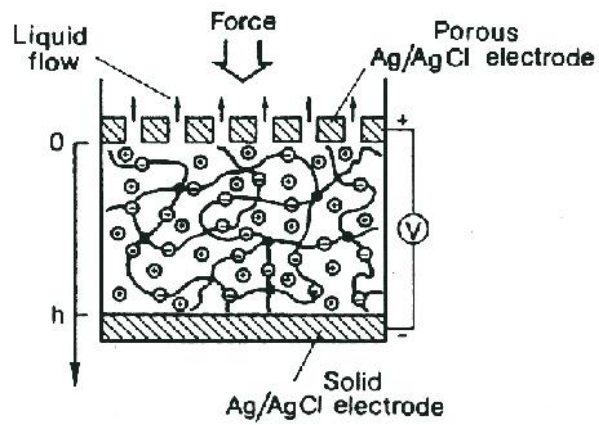

*Figure 16* Schematic representation of the streaming potential operating in a sample of a negatively charged polymer network with positive mobile counter-ions. When the sample is compressed, the generated water flow through the porous electrode induces the mobile charge displacement that originates an electric potential.

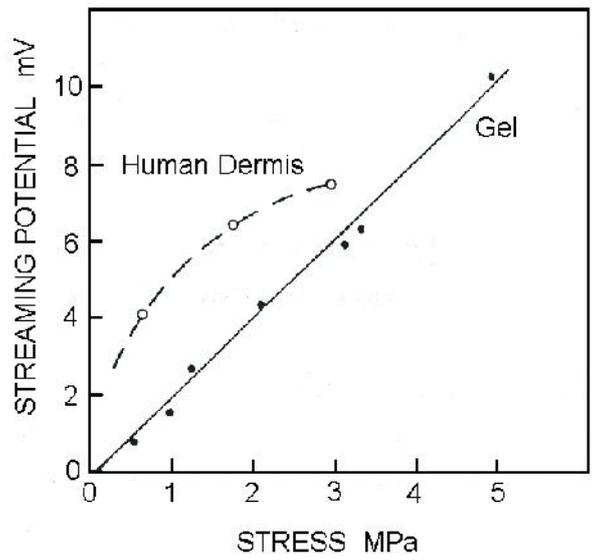

*Figure 17* *Stress-generated potentials versus applied load for 400 μm thick polyvinylalchol-polyacrylic-acid gel and Human skin samples in water.*